# Securing Biometric Data: Fully Homomorphic Encryption in Multimodal Iris and Face Recognition

Surendra Singh, Lambert Igene, Stephanie Schuckers [1]

**Abstract:**

Multimodal biometric systems have gained popularity for their enhanced recognition accuracy and resistance to attacks like spoofing. This research explores methods for fusing iris and face feature vectors and implements robust security measures to protect fused databases and conduct matching operations on encrypted templates using fully homomorphic encryption (FHE). Evaluations on the QFIRE-I database demonstrate that our method effectively balances user privacy and accuracy while maintaining a high level of precision. Through experimentation, we demonstrate the effectiveness of employing FHE for template protection and matching within the encrypted domain, achieving notable results: a 96.41% True Acceptance Rate (TAR) for iris recognition, 81.19% TAR for face recognition, 98.81% TAR for iris fusion (left and right), and achieving a 100% TAR at 0.1% false acceptance rate (FAR) for face and iris fusion. The application of FHE presents a promising solution for ensuring accurate template matching while safeguarding user privacy and mitigating information leakage.

## 1 Introduction

Biometric authentication systems have become increasingly widespread in enhancing security protocols across various sectors due to their accuracy and reliability compared to traditional methods. Among these systems, multimodal biometric authentication, which combines multiple biometric modalities such as fingerprints, facial recognition, and iris scans, has proven to be effective for cases when enhanced security is needed. This approach harnesses the strengths of each modality to offset the weaknesses of others, thereby providing a more comprehensive and secure authentication process. Research indicates that multimodal biometric systems not only improve the accuracy of identity verification but also enhance resistance to spoofing attacks and other fraudulent activities [JR04]. By integrating various biometric traits, multimodal systems ensure higher reliability and security, making them an essential component in the advancement of authentication technologies.

To ensure integrity and facilitate widespread adoption, biometric recognition systems must address several critical security concerns [JRU05] some of these include: *privacy concerns, data breaches* and *misuse of biometric data*. Protecting the biometric template database is paramount to maintaining the system's accuracy and security [MPR20]. This study introduces an efficient multimodal recognition system that combines iris and face biometrics, enhanced with fully homomorphic encryption (FHE) to bolster security. Homomorphic encryption aligns with the standards outlined in ISO/IEC 24745 [Pa22], which provides comprehensive guidelines for protecting biometric information. These guidelines cover key aspects such as maintaining confidentiality, ensuring data integrity, and supporting the renewability and revocability of biometric templates during both storage and transmission. The recognition process uses encrypted versions of the fused iris and face templates, significantly improving the system's security. Unlike traditional encryption methods, FHE allows computations to be performed directly on encrypted data without the need to decrypt the templates for matching. This capability significantly enhances the security of the biometric system while maintaining the accuracy of the recognition process. To optimize efficiency and manage the size of the final fused template, principal component analysis (PCA) is applied to reduce the dimensionality of the iris data. We evaluate the system's accuracy by measuring performance across various face recognition models, recognition of a single iris, and fusion of both irises. This approach ensures a robust, secure, and efficient biometric authentication system.

---

[1] Department of Electrical and Computer Engineering, 8 Clarkson Ave, Potsdam, NY,
{sursing, igenela, sschucke} @clarkson.edu

Template protection techniques for biometric data can be classified into two primary categories: feature transformation and biometric cryptosystems [Hu22]. Biometric cryptosystems involve techniques like key binding and key generation to secure the data [BK12, JRU05]. However, these methods often degrade recognition accuracy due to the inherent variability in biometric data [NJ15]. Homomorphic encryption (HE) presents a promising solution for protecting biometric data [Mi23]. HE enables specific algebraic operations to be performed on ciphertext-encrypted data such that the decrypted result matches the outcome of performing the same operations on the plaintext [Ar15, JZR17, Ma21, Wu21]. This capability ensures that biometric templates can be securely stored and processed without compromising their privacy. The primary benefits of HE in biometric systems are enhanced privacy, security, and flexibility. It allows secure storage and computation on encrypted biometric templates, maintaining data privacy during matching and reducing the risk of unauthorized access, thereby strengthening system security [ATBS15]. This study focuses on developing a multimodal privacy-preserving authentication approach using fully homomorphic encryption (FHE), addressing template security and computational efficiency. The proposed system integrates iris and face recognition models enhanced by FHE. Initially, the face recognition model captures the final face feature vector, while PCA reduces the dimensionality of fused left and right iris codes, preserving essential features for improved system efficiency and accuracy. These encrypted feature vectors are fused and evaluated using the Euclidean distance metric on the QFIRE-I [Jo10] database, ensuring robust performance across various real-world conditions.

## 2 Related work on homomorphic encryption for biometrics

Sperling *et al.* [Sp22], proposed a method to fuse face and voice biometric templates, directly in the encrypted domain through a linear projection followed by normalization. Vallabhadas *et al.* [VS22], proposed a privacy-preserving authentication system that en-hances accuracy and reduces computation time. The system transforms original templates into pseudo templates by fusing iris and fingerprint traits. It reduces the iris template size using block size reduction, then applies Local Random Projection (LRP) for revocable and unlinkable templates, and uses FHE for privacy. Tested on the Children Multimodal Biometric Database (CMBD), it achieved an Equal Error Rate (EER) of 0.0214%. Sharma *et al.* [Sh23], examines the potential for implementing score-level and decision-level fusion using FHE to improve security and privacy. Morampudi *et al.* [MSD23], proposed a secure authentication system with fuse iris and fingerprint templates. They proposed an optimized method for calculating the Hamming distance between encrypted templates. Template dimension reduction achieved by $d$-sized block reduction technique. This method achieved 0.16% EER on CMBD and 0.24% EER on IITD database with a template dimension of 4018.

Numerous studies have explored the application of FHE in face recognition systems, with a primary focus on safeguarding users' privacy [Ya22, MZ22, Ar23, Bo18]. Morampudi *et al.* [MPR20], proposed a privacy-preserving iris authentication method using FHE. The proposed method calculates the Hamming distance of the encrypted iris templates for matching. Morampudi *et al.*[Mo21], proposed SvaS, a secure and verifiable classification-based iris authentication system using FHE. Block size reduction compresses the iris templates, and matching is performed on encrypted data using NN (nearest neighbor) and MCP (multi-class perception). Bassit *et al.*[Ba23a, Ba22], proposed multiplication-free methods for matching biometric templates. Engelsma *et al.* [EJB22], proposed Deep-MDS++ method to reduce the template dimension, and this method was demonstrated on the face and fingerprint templates. Bauspieß *et al.* [Ba23b], propose homomorphic transciphering to enhance security against offline decryption attacks without compromising biometric performance, enabling multi-biometric comparisons of fixed-length feature presentations.

In this research, we propose a secure fusion of iris and face templates using FHE, enabling secure matching on encrypted templates. To enhance efficiency, we incorporate PCA to reduce the dimensionality of the iris features while retaining the most informative components. These reduced dimension iris features are then fused with the face feature vector, resulting in a more compact and efficient representation for secure matching. Unlike previous research, which reduces the size of the iris fusion template through simplistic pixel-skipping approaches, our method employs PCA in the reduction process. PCA effectively retains crucial features, whereas skipping bits or pixels fails to preserve meaningful data patterns, potentially resulting in information loss and suboptimal representations. This enhancement not only optimizes template size but also enhances the robustness and efficiency of fusion recognition systems.

## 3 Homomorphic encryption encrypted template and keys details

FHE implementation consists of multiple components, each serving a distinct function within the encryption process. Table 1 shows the details of the encrypted template and key size details. *Public key:* The public key encrypts the data, which is then sent to a third party, such as a cloud server, for computation on the encrypted data. *Secret key:* The secret key decrypts the computation results once they are sent back to the client device.

|  |  | Template/key size |
|---|---|---|
| Key | Public key | 736 KB |
|  | Secret key | 272 KB |
| Encrypted template | Enrollment encrypted template | 128 KB |
|  | Verification encrypted template | 128 KB |

Tab. 1: Key sizes and encrypted template sizes for Fully Homomorphic Encryption.

### 3.1 Architecture flow for enrollment and verification processes and secure matching

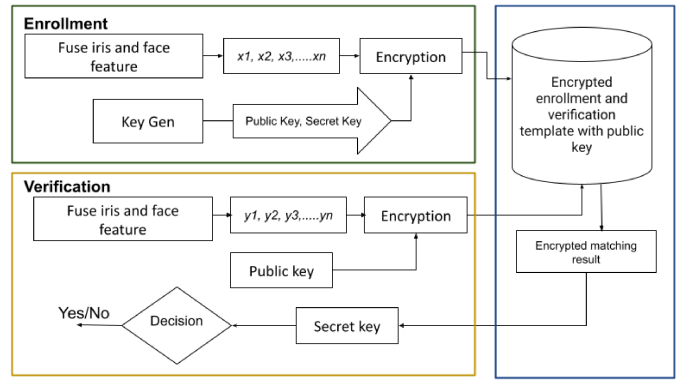

Fig. 1: Enrollment and verification processes for iris and face fusion matching with FHE.

We encrypt fuse templates using FHE and conduct matching directly in the encrypted domain. Figure 1 illustrates the recognition process. In the enrollment and verification processes, the architecture involves the FR model for generating face features and OSIRIS [ODGS16] for iris features. Key generation (Keygen) produces the public and secret keys. The encrypted template, created with the public key, is sent to the cloud server with the public key. The cloud server performs the matching process and returns the encrypted matching result to the verification device. The secret key is then used to decrypt the matching score.

**Secure matching:** In our research, we perform one-to-one matching. We implemented the homomorphic Euclidean distance to measure the matching score between the gallery and probe encrypted template. Algorithm 1 provides a step-by-step description of the mentioned procedure. The algorithm involves two encrypted vectors, a gallery vector $x$, and a probe vector $y$, and returns the encrypted distance $dt$ which can be decrypted by using *secretkey*. We also implemented the matching algorithm proposed by Boddeti [Bo18], which utilizes the inner product between two vectors. This matching algorithm requires three sets of keys: Public, Galois, and Relin keys. Notably, the Galois key itself is 9.2MB in size, while the Relin key is 408.5KB. In contrast, our proposed homomorphic Euclidean distance matching approach only necessitates a pair of public and private keys. Table 4 illustrates a comparative analysis of the matching scores between the two different algorithms, highlighting the accuracy and practicality of our proposed method.

**Algorithm 1** Homomorphic Euclidean distance

```
 1: procedure ENCRYPTED EUCLIDEAN DISTANCE(x, y)
 2:     ε(x) ← f(x, θe)                              ▷ Gallery Encoding
 3:     ε(y) ← f(y, θe)                              ▷ Probe Encoding
 4:     st ← Subtract(ε(x), ε(y))                    ▷ Subtract
 5:     dt ← DotProduct(ε(st), ε(st))                ▷ Dot Product
 6:     return dt                                    ▷ Return the encrypted distance
 7: end procedure
```

## 4 Database Overview

In this research, we use Q-FIRE [Jo10] database. Each subject visit includes 2 left and 2 right iris images and 10 face images. The face dataset has 1804 total images with 155 subjects. The split between the gallery and probe data is 70% face images consumed for the gallery and 30% face images for probe data. The iris dataset has 119 total subjects with 476 total images. Each subject has 2 images for left and right iris images in two different sessions. Figure 2 example of iris and face images. This database include face images from multiple angles (frontal, profile, and oblique) to account for the diversity in real-world scenarios. Additionally, the database must incorporate images with motion blur, which simulates the effects of movement, and out-of-focus blur, representing the challenges posed by varying camera focuses. By including these variations, the database can better support the development and testing of biometric systems that are resilient to common issues in facial recognition, ultimately enhancing their reliability and security. Table 2 explain the database details.

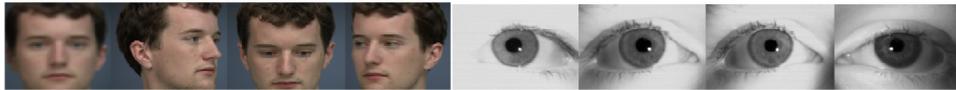

Fig. 2: Example of face and iris images from the QFIRE-I database which includes different angles and levels of focus.

## 5 Accuracy analysis of individual iris and face recognition

To conduct the fusion of iris and face data, we first run experiments on each modality separately to establish a comparison benchmark. In the following sections, we detail the approach and model selected for face recognition (FR) and the dimensionality reduction of iris code.

**Face detection and recognition models:** We used RetinaFace [De19] for face detection to accurately detect and align faces. In this study, we first compare ArcFace, Deepface, Facnet512, Facnet, OpenFace and MagFace for feature extraction. Table 3 provides the final feature vector length for each face matcher and accuracy details across all the face matchers. We noticed that MagFace performance is comparably better performance with 81.19% TAR at 0.1% FAR for the low quality images of the QFIRE database. In this research, we aim to reduce the final iris code dimension while generating a balanced and fused iris and face feature vector length. To achieve this, we employ a DNN face recognition model that ensures the feature vector is minimized in length without compromising accuracy. It is essential to manage the feature vector size because if it becomes too large, the FHE computation can become unwieldy. This results in extended processing times, which can negatively impact the efficiency and practicality of the biometric system. Thus, our approach not only focuses on feature reduction but also on optimizing computational efficiency.

|  | Distance (ft) | Description |
|---|---|---|
| **Iris** | | |
| Out-of-focus blur | 5, 7, 11, 15, 25 | Full range of blur |
| Illumination | 5, 7, 11, 15, 25 | Low, Med, High |
| Angles | 5, 7, 11 | Straight, Left, Right, Up, Down |
| Occlusion | 5, 7, 11, 15, 25 | 6 seconds of blinking |
| Motion blur | 7, 15 | Slow/Fast Walking |
| **Face** | | |
| Out-of-focus blur | 5, 7, 11, 15, 25 | Full range of blur |
| Illumination | 5, 7, 11, 15, 25 | Low, Med, High |
| Angles | 5, 7, 11 | Straight, Left, Right, Up, Down |
| Occlusion | 5, 7, 11, 15, 25 | Multiple Faces |
| Motion blur | 7, 15 | Slow/Fast Walking |

Tab. 2: QFIRE [Jo10] database details.

| Model | Feature length | TAR@0.01% FAR | TAR @0.1% FAR | TAR @1% FAR |
|---|---|---|---|---|
| ArcFace | 512 | 26.89 | 31.91 | 40.90 |
| DeepFace | 160 | 27.12 | 29.43 | 34.35 |
| Facenet | 128 | 36.2 | 50.69 | 65.91 |
| Facenet512 | 512 | 32.51 | 46.43 | 57.15 |
| OpenFace | 128 | 29.84 | 42.40 | 60.85 |
| MagFace | 512 | **71.34** | **81.19** | **88.67** |

Tab. 3: Face recognition accuracy with different FR matcher with FHE.

**Iris recognition:** We performed iris recognition to match the left and right irises of individuals. In this research, we used the OSIRIS [ODGS16] tool for feature extraction. The final iris code dimension is $128 \times 512$, and to avoid rotational inconsistency, we applied ±7 rotation shifts and stored all 15 templates in the gallery database. We transformed the final iris code matrix with dimensions $128 \times 512$ into a vector of size $65,536 \times 1$ by applying a row-wise flattening operation during the vectorization process. We then applied PCA to reduce the final iris code dimension to 250 feature length. We achieved a 95.89% TAR at 0.01% FAR and a 96.41% TAR at 0.1% FAR.

## 6 Iris and Face fusion

In this study, we explored approaches to fuse iris and face templates to enhance biometric system performance. Initially, we fused the left and right iris templates, followed by their integration with the final face feature. The subsequent sections elaborate on these fusion techniques and their implementation.

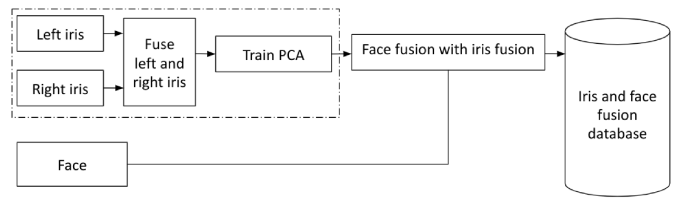

Fig. 3: Approach: Left and right iris fusion, followed by PCA to reduce the iris dimension, and finally fusion with the face.

**Iris database creation:** We joined the left and right iris codes together. The figure 3 represents the final fuse iris code. In the first step, we concatenate the left and right iris code and after that perform ±7 rotations to avoid rotation inconsistency. To apply PCA for dimension reduction, we randomly selected 50% of gallery data to train the PCA and save this train PCA variance to apply on gallery and probe data to reduce the dimension. We matching one probe with 15 gallery templates, one original and 14 with rotational shifts, and select the minimum score amongst all the templates for accuracy measurement.

**Iris and face fusion database:** We use MagFace face matching for final feature extraction, which generates a 512 dimensional feature vector. After fusing the left and right iris templates and applying PCA to reduce the final dimension, we use a 500-dimensional feature vector for both the left and right iris. Subsequently, after fusing both the face and iris final features, the dimension becomes 1012. We split the face data, allocating 30% for the probe and 70% for the gallery dataset. We pair all gallery face images with each subject gallery's final iris images. The fused iris code dimension reduces from 131072 to 500, while the face feature vector length remains at 512, resulting in a final fused iris and face feature vector length of 1012. Figure 3 illustrates the data flow to create the final iris and face fused template. The results, as detailed in Table 4, provide the accuracy details. We achieved 100% TAR at 0.1% and 0.01% FAR for iris and face fusion at a 1012 feature vector length. We also compared our results with the matching algorithm proposed by Boddeti [Bo18]. In addition to encrypted vectors, we also compared Euclidean distance with plain text.

## 7 Discussion

**Computational Efficiency:** To evaluate the computational efficiency, we compared one-to-one matching (one probe and one gallery) with and without encryption. We found that with FHE, it takes 1.05 seconds, while with plain text, it takes 0.63 seconds. The tests were conducted on a machine with $8-coreIntel(R)Core(TM)i7-10700CPU$ at $2.90GHz$.

|  | | Euclidean distance | | Inner product [Bo18] | | Plain text | |
|---|---|---|---|---|---|---|---|
| Fusion | Feature length | TAR @0.1% FAR | TAR @0.01% FAR | TAR @0.1% FAR | TAR @0.01% FAR | TAR @0.1% FAR | TAR @0.01% FAR |
| Face only | 512 | 71.34 | 81.19 | 70.53 | 79.83 | 71.34 | 81.19 |
| Single iris only | 250 | 96.41 | 95.89 | 96.41 | 95.89 | 96.41 | 95.89 |
| Fusion: left and right iris | 500 | 98.81 | 97.62 | 95.0 | 94.0 | 98.81 | 97.62 |
| Fusion: Left and right joint iris PCA, Face | 1012 | **100** | **100** | **100** | **100** | 100 | 100 |

Tab. 4: Accuracy analysis of iris, face, iris fusion, and iris with face fusion.

**Score comparison:** To compare individual modality scores, we selected a face subject with a match score greater than the threshold 0.4. The average score among all gallery matches was 0.7, indicating that this subject would be considered a false rejection. In contrast, for the iris modality, the matching score was 0.8, while the iris matching threshold was 0.84. This is classified as a true match at the border threshold. The fusion score for the iris was 0.49, with a fusion threshold of 0.61, indicating that the fusion resulted in a true acceptance. Finally, with both iris and face fusion, the score was 0.85, compared to a threshold of 0.87. These results demonstrate that fusing face features with iris data improves system accuracy compared to using the face modality alone.

## 8 Conclusions

In this study, we have investigated a method to fuse iris and face feature vectors that incorporates robust security measures to safeguard databases and conduct matching operations on encrypted templates using FHE. Our findings demonstrate the efficacy of employing FHE for template protection and matching directly within the encrypted domain, yielding impressive results: a 96.41% TAR for iris alone, 81.19% TAR for face recognition, 98.81% TAR for iris fusion (left and right), and 100% TAR at 0.1% FAR for face and iris fusion. The application of FHE offers a promising solution for accurate template matching while preserving user privacy and preventing information leakage. Our research demonstrates the feasibility of integrating FHE into multimodal biometric template protection protocols. Future research should explore deep learning methodologies to minimize the dimensionality of iris codes, enhancing biometric authentication systems' efficiency and efficacy. Additionally, optimizing the size of encryption keys and templates is imperative in minimizing data transfer times, thereby enhancing operational efficiency and ensuring seamless integration into existing systems. This strategic approach not only bolsters security but also promotes the practical implementation of FHE in diverse technological landscapes. Future work should also evaluate the performance of FHE on large databases for efficient identification.